\documentclass[useAMS,usenatbib]{mn2e}
\usepackage{natbib}
\usepackage{graphicx}
\usepackage{floatflt}
\usepackage{amssymb}
\newcommand{\comment}[1]{}
\newcommand{\lam}{$\lambda$}

\newcommand{\be}{$\beta$ }
\newcommand{\fhils}{\textit{FHILs}}
\newcommand{\fhil}{\textit{FHIL}}
\newcommand{\kms}{km s$^{-1}$}
\newcommand{\e}[1]{$\times 10^{#1}$}
\title[Models of FHIL emission]{The location and kinematics of the coronal-line emitting regions in AGN}
\author[J. R. Mullaney, M. J. Ward, C. Done, G. J. Ferland, \& N. Schurch]{J. R. Mullaney$^{1}$\thanks{E-mail: j.r.mullaney@dur.ac.uk},  M. J. Ward$^{1}$, C. Done$^{1}$, G. J. Ferland$^{2,3}$, \& N. Schurch$^{4}$  \\
$^{1}$Department of Physics and Astronomy, Durham University, South Road, Durham, DH1 3LE, U.K.\\
$^{2}$Institute of Astronomy, University of Cambridge, Madingley Road, Cambridge, CB3 0HA, U.K.\\
$^{3}$Department of Physics, University of Kentucky, Lexington, KY 40506, USA\\
$^{4}$Key Laboratory for Particle Astrophysics, Institute of High Energy Physics, Chinese Academy of Science, Beijing 100049 China} 
\begin{document}

\date{Date Accepted}

\pagerange{\pageref{firstpage}--\pageref{lastpage}} \pubyear{2007}

\maketitle

\label{firstpage}

\begin{abstract}
We use the photoionisation code \textsc{cloudy} to determine both the location and the kinematics of the optical forbidden, high ionisation line (hereafter, \fhil) emitting gas in the narrow line Seyfert 1 galaxy Ark 564.  The results of our models are compared with the observed properties of these emission lines to produce a physical model that is used to explain both the kinematics and the source of this gas.  The main features of this model are that the \fhil \ emitting gas is launched from the putative dusty torus and is quickly accelerated to its terminal velocity of a few hundred km s$^{-1}$.  Iron-carrying grains are destroyed during this initial acceleration.  This velocity is maintained by a balance between radiative forces and gravity in this super-Eddington source.  Eventually the outflow is slowed at large radii by the gravitational forces of and interactions with the host galaxy.  In this model, \fhil \ emission traces the transition between the AGN and bulge zones of influence.
\end{abstract}


\section{Introduction}
The presence of \fhils \ in the spectra of AGN is a signature of the high energy processes occurring in these objects.  It has been argued that these lines are a direct response to the strong ionising continuum of AGN as they are produced by high ionisation states with potentials $>$100 eV, although collisional excitation is still not ruled out as a possible production mechanism (see \citealt{Penston84}).
  
Early studies of \fhils \ (e.g. \citealt{Osterbrock81}, \citealt{Penston84}, \citealt{DeRobertis84}) revealed that their profiles differed from the more prominent low ionisation forbidden lines (such as [O III]\lam5007 and [N II]\lam6584) \textit{and} the permitted broad lines (e.g. H\be and H$\alpha$).  In addition to having widths between these two extremes, they were often observed to be blueshifted with respect to the low ionisation lines or to possess an asymmetric blue wing, suggesting a net flow of the gas producing these lines.  Irrespective of the interpretation of the source of these kinematics, it is clear that at least part of the \fhil \ emission arises from a region of the AGN distinct from the traditional broad line and narrow line regions (BLR and NLR, respectively; see \citealt{Mullaney08} for a more detailed discussion).  Further evidence that \fhils \ arise from an `intermediate' region (i.e. between the BLR and NLR) is provided by integral field observations of the near infrared \fhils \ in two of the closest AGN: the Circinus galaxy and \textit{NGC 4051} (\citealt{Mueller06} and \citealt{Riffel08}, respectively).  These studies showed that these \fhils \ were being emitted at up to 75 pc from the nucleus.

In their study of the \fhils, \cite{Ferguson97} (hereafter F97) used photoionisation models to show that a typical AGN spectral energy distribution (hereafter SED) could produce these lines in clouds covering a wide range of distances from the central engine; from just outside the BLR to the extended NLR.  However, as their study focussed on the \textit{relative} positions of the emitting regions of various \fhils \ it did not address the issue of the source of these clouds and, in turn, explain the kinematics of these clouds evident in the profiles of the lines. 

Here, we extend the work of F97 by presenting the results of a new set of photoionisation models that, when combined with the observed properties of the \fhils, produce a self-consistent picture of the source and kinematics of the \fhil \ emitting clouds around AGN. \cite{Dorodnitsyn08} recently used hydrodynamic models to show that a wind launched from the inner edge of an X-ray illuminated torus would reach the velocities observed in the absorption lines of the so-called warm absorber ($\lesssim$500 km s$^{-1}$ to 1000 km s$^{-1}$).  As the velocity shifts associated with this feature are similar to those of the \fhils \ we investigate whether this same wind could be responsible for the observed shifts of these lines.

Throughout this \textit{Letter} we use the convention that [Fe VII] refers to the [Fe VII]\lam6087 line (ionisation potenial: 100 eV), [O III] to [O III]\lam5007 (35 eV), [Fe X] to [Fe X]\lam6374 (235 eV) and [Fe XI] to [Fe XI]\lam7892 (262 eV).

\begin{figure}
\begin{center}
	\includegraphics[width=8.0cm, height=6.3cm]{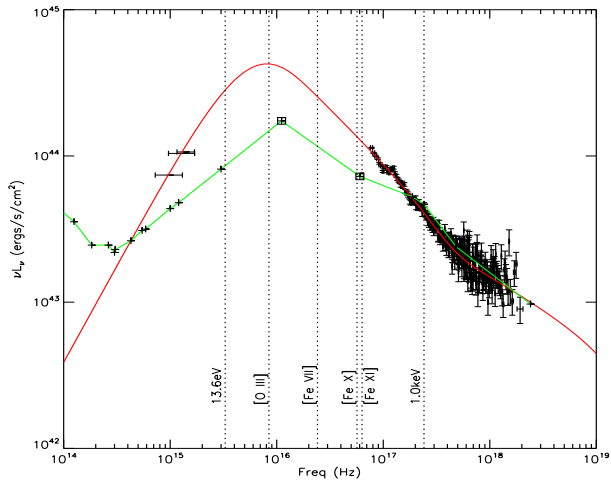}
\end{center}
\caption{The two SEDs used in our models.  The green line represents SED-R04 of \protect \cite{Romano04}.  Points surrounded by squares have been extrapolated from the observable portions of the SED. The red line represents SED-X constructed using optical/UV/X-ray data obtained with the XMM telescope.  Dotted lines indicate the ionisation potentials of the species discussed here.  Lines showing the 13.6 eV and 1 keV are included for reference.}
\label{SEDs}
\end{figure}

\section{The models}
\label{Models}
We base our models on the observed emission lines and continuum properties of the NLS1, Ark 564,  as it is known to have very strong \fhils \ (e.g. \citealt{Erkens97}, \citealt{Mullaney08}, \citealt{Riffel06}, \citealt{Landt08}).
Calculations were performed with version 08.00 of \textsc{cloudy}, last described by \cite{Ferland98}, to predict the emission lines shown in the optical spectrum of Ark 564 taken with the \textit{ISIS} spectrometer on the WHT telescope (see \citealt{Mullaney08} for details regarding these observations). 

\subsection{Input SED}
\label{SED}
In their study of the low ionisation lines of Ark 564 \cite{Romano04} assembled an SED (hereafter SED-R04) that was constrained by their observations between the mid-IR$-$UV and 1$-$10 keV.  They extrapolated the UV and soft X-ray continuum to form an SED that peaks at $\sim$40 eV in $\nu F_\nu$ space (see fig. \ref{SEDs}).  This was used as our initial input SED.  To determine the sensitivity of our results to the shape of the strongly absorbed portion of the SED, we assembled our own based on XMM observations of Ark 564  taken on the 17$^{\mathrm{th}}$ June, 2000 (hereafter SED-X). This SED peaks at approximately the same photon energy, but has a luminosity at this energy roughly twice that of SED-R04.  Because the major difference between the two SEDs occurs within the unobservable section of the electromagnetic continuum, SED-X conforms to the observed fluxes in the optical and X-ray wavebands.  It should be noted, however, that SED-X overproduces flux at 1000 \AA \ (3\e{15} Hz) as compared to SED-R04 by a factor of $\sim$2, although this could be explained in terms of increased reddening which is severe at these wavelengths.  We normalised both SEDs to have a luminosity of $2.64\times10^{43}$ ergs s$^{-1}$ at 7000\AA \ (4.3\e{14} Hz; \citealt{Romano04}) resulting in total ionising luminosities of 4.8\e{44} ergs s$^{-1}$ and 1.1\e{45} ergs s$^{-1}$ for SED-R04 and SED-X, respectively.  This corresponds to Eddington ratios of $L/L_{Edd}=1.8$ and 3.9 (respectively) assuming a black hole mass of 2.6\e{6} M$_{\odot}$ (\citealt{Botte04}). 

\begin{figure}
\begin{center}
	\includegraphics[width=8.0cm, height=6.3cm]{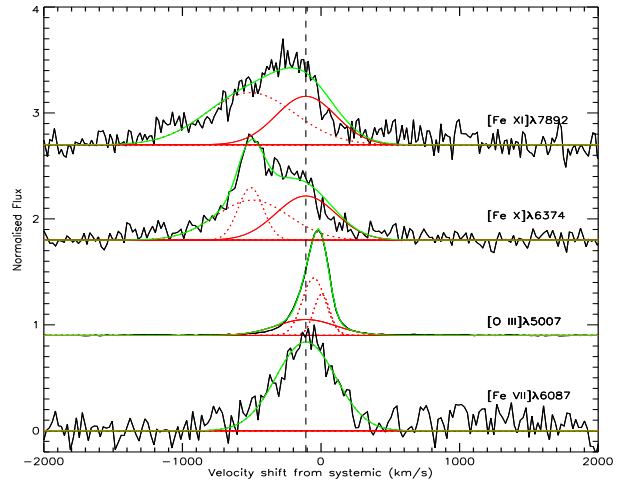}
\end{center}
\caption{[O III], [Fe X] and [Fe XI] profiles decomposed into gaussian components, one of which has the same kinematic parameters as the fit to the [Fe VII] line.  This places upper limits on the flux produced with the same kinematics as the [Fe VII] line.}
\label{Profiles}
\end{figure}

\begin{table}
\centering
\caption{Component parameters for the emission lines.  The [Fe VII] line was fit with a single component.  The other line fits were then forced to contain a component with the same FWHM and velocity shift.} 
\label{Components}
\begin{tabular}{lcccc}
\hline
FHIL&Flux$^a$&Shift$^b$ (km s$^{-1}$)&FWHM (km s$^{-1}$)\\
\hline
$[$O III$]$&6.2&-110&507\\
\hspace{0.2cm}&4.3$\pm$0.3&6$\pm$1&133$\pm$2\\
\hspace{0.2cm}&8.6$\pm$0.3&-52$\pm$2&193$\pm$1\\
$[$Fe VII$]$&1.00$\pm$0.03&-110$\pm$3&507$\pm$7\\
$[$Fe X$]$&1&-110&507\\
\hspace{0.2cm}&1.1$\pm$0.1&-494$\pm$13&620$\pm$30\\
$[$Fe XI$]$&0.7&-110&507\\
\hspace{0.2cm}&1.1$\pm$0.1&-510$\pm$12&750$\pm$28\\
\hline
\end{tabular}
\begin{flushleft}
$^a$Component flux relative to [Fe VII].\\
$^b$Velocity shift; negative values are blueshifts.
\end{flushleft}
\end{table}

\subsection{Cloud geometry, density and composition}
\label{Geometry}
In their work on the line emitting regions of AGN F97 showed that \fhil \ emission peaks along ridges of constant ionisation paramater, $U$ ($=L_{ion}/4\pi R^2n_Hh\nu c$; where $R$ is the distance of the emitting cloud of hydrogen density $n_H$ from a continuum source with an ionising luminosity of $L_{ion}$; all other symbols have standard meanings).  More specifically they noted that the optical iron \fhils \ all reached a maximum luminosity around $R\sim10^{18}-10^{19}$ cm (0.3 $-$ 3 pc) and $n_H\sim10^5-10^7$ cm$^{-3}$ for a source of ionising luminosity $L_{ion}=5\times10^{43}$ ergs s$^{-1}$.  These values act as a rough guide to the location of the \fhil \ emitters in Ark 564.  However, as we include radiative driving in our models which tends to reduce the density of the emitting clouds and the ionising luminosity is greater in Ark 564 we chose to model a large portion of the parameter space covering $15<log(R)<20.5$ and $4<log(n_H)<11$ (in steps of 0.5 for both $R$ and $n_H$) to ensure we cover all the possible \fhil \ emitting regions.

We have modelled our clouds as being dust free.  F97 showed that the inclusion of dust in their models caused the \fhils \ to be suppressed by up to a factor of 25 compared to the dust free models.  This is largely the result of absorption of the ionising radiation by the dust grains as well as the depletion of gas phase Fe onto the dust grains.  Finally, we assumed solar metallicities (our exact ratios are the same as those given in F97)

\begin{figure}
\begin{center}
	\includegraphics[width=8.4cm, height=7.5cm]{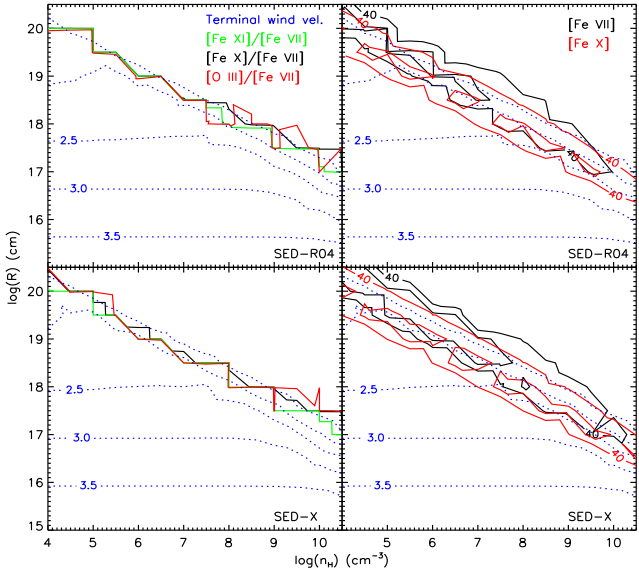}
\end{center}
\caption{Contour plot showing the regions of the parameter space at which the observed $k$-ratios (see \S\ref{Dynamics}) are produced (left column) and the [Fe VII] and [Fe X] lines are most strongly emitted (right column).  The dotted blue contours show the log(terminal velocity/\kms) of the clouds at each point in the parameter space.  An electronic version of this plot is available at www.dur.ac.uk/j.r.mullaney.}
\label{Contours}
\end{figure}

\subsection{Cloud dynamics}
\label{Dynamics}
Any model designed to reproduce the observed line ratios should also be able to reproduce the velocity structure of the \fhil \ emitting region.  To account for this we included radiative driving in our models by using the `wind' command in \textsc{cloudy}, the physics of which is described in \cite{Henney05}.  In allowing the clouds to be radiatively driven, we drastically change the physical parameters of the emitting gas as compared to models involving only stationary clouds.  This has corresponding effects on the modelled line ratios.  

By conservation of mass, the density and velocity of a continuous column of gas is related such that $4\pi r_0^2\rho_0v_0=4\pi r_1^2\rho_1v_1$, where $\rho_{0,1}$ and $v_{0,1}$ represent the density and velocity at two radii, $r_{0,1}$.  In including radiative driving we therefore decrease the density of the emitting clouds as a function of depth, allowing ionising radiation to penetrate deeper into the cloud.

The profiles of the emission lines were used to further constrain our models.  By decomposing the profiles of the iron \fhils \ and [O III] into gaussian components we can place upper limits on the luminosity of the lines emitted from a region described by a single set of kinematic properties.  To do this we fit the [Fe VII] line with a single gaussian (a second component was found to be statistically unnecessary) and force the fits of the other emission lines to contain a component with the same velocity shift and FWHM (507 $\pm$ 7 km s$^{-1}$ and -110 $\pm$ 3 km s$^{-1}$, respectively; see fig. \ref{Profiles} and table \ref{Components}).  From this analysis we note that our models of the [Fe VII] emitting gas must not produce [O III] at luminosities greater than $\sim$6 times that of [Fe VII], while the [Fe X]/[Fe VII] and [Fe XI]/[Fe VII] ratios of kinematically similar components (hereafter referred to as $k$-ratios) should be no more than unity and 0.61, respectively.  To account for this and to reduce computation time our model calculations were stopped when [O III]/[Fe VII]$\geq$6.  Furthermore, we note that fig. \ref{Profiles} shows little or no [Fe X] or [Fe XI] flux being produced at blueshifts lower than the [Fe VII] component.

Because photons that are most capable of producing Fe$^{6+}$ to Fe$^{10+}$ are readily absorbed, \fhils \ will be produced nearer the front face of the clouds (i.e. facing the continuum source) while the lower ionisation lines will be preferentially produced at larger depths.  By placing an upper limit on the flux of the [O III] emission from the \fhil \ emitting cloud we can therefore limit the size of the modelled cloud.  

Other important stopping conditions were included to prevent the clouds from becoming geometrically thick or optically thick ($\Delta R/R < 1$ and $N_H < 10^{24}$ cm$^{-2}$).  Following F97 the calculation was stopped if the temperature of the cloud dropped below 3000K to prevent fragmentation.  In practice the [O III]:[Fe VII] ratio always became $>$6 before this stopping criterion was met.

\section{Results}
\label{Results}
\begin{figure}
\begin{center}
	\includegraphics[width=8.0cm, height=4.1cm]{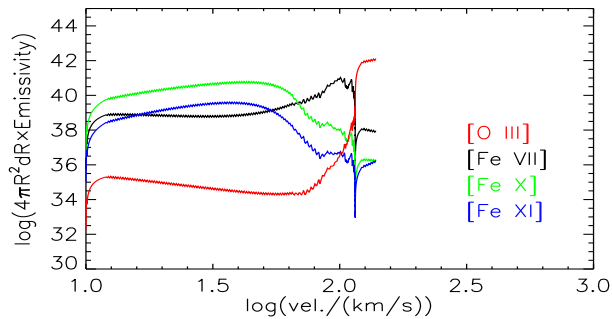}
\end{center}
\caption{Emission line luminosity as a function of gas velocity for a cloud at $R=10^{19}$ cm with $n_H=10^6$ cm$^{-3}$ illuminated by SED-X ($\textrm{log}(U)=-0.4$)}.
\label{EmVel_19_6}
\end{figure}

\begin{figure}
\begin{center}
	\includegraphics[width=8.0cm, height=4.1cm]{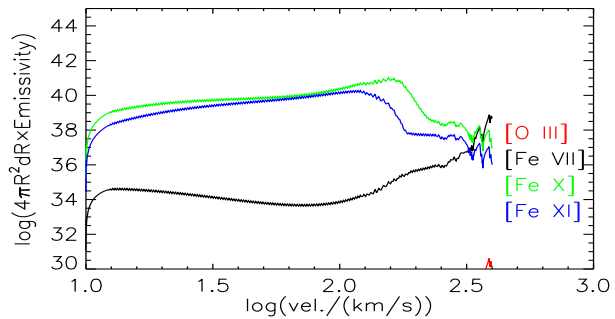}
\end{center}
\caption{Same as fig. \ref{EmVel_19_6} but for clouds at $R=10^{17}$ cm and $n_H=3\times10^9$ cm$^{-3}$ ($\textrm{log}(U)=0.1$)}.
\label{EmVel_17_95}
\end{figure}

In fig. \ref{Contours} we present contour plots showing the regions of the parameter space at which SED-R04 and SED-X produce the observed $k$-ratios (defined in \S \ref{Dynamics}) and the [Fe VII] and [Fe X] lines are most strongly produced.  The [Fe XI] contours lie close to the [Fe X] contours and have been removed for clarity.  Also shown on these plots is the velocity the modelled clouds reached by the end of the calculation.  This could represent either the true terminal velocity of the cloud or the velocity that the cloud reached at the point when the calculation was halted, e.g. when the [O III]/[Fe VII] $k$-ratio reached 6 (these two values may not be the same).  Furthermore, the \fhils \ are not necessarily strongest at the terminal velocity- in general they peak at lower velocities (see figs. \ref{EmVel_19_6} and \ref{EmVel_17_95}).  The velocity characteristics of fig. \ref{Contours} are a result of the fact that when the clouds are optically thick essentially all the incident photon momentum is transferred to the gas.  At higher cloud densities the total mass of the clouds will increase and hence their terminal velocities decrease. 

Both SEDs are capable of producing the observed $k$-ratios along a line of constant log($U)=-0.4$ stretching from log($n_H$)=5.0, log($R$)=19.5 to log($n_H$)=10, log($R$)=17.0.  This suggests that when kinematics are included in the models, the \textit{precise} shape of the illuminating SED has little effect on these ratios.  However, since both SEDs were produced to match the observed emission of a single object, this does not rule out the possibility that \fhil \ ratios will be affected by the large differences in overall SED shape between different AGN.  The influence of the SED shape on the relative strengths of the \fhils \ will be the focus of a later paper.

Neither SED is capable of (radiatively) driving the emitting clouds to the observed blueshifts of the \fhils \ in the regions of the parameter space where these lines are most strongly produced.  Even the increased EUV flux of SED-X can only produce velocities of $\sim$30 \kms \ and $\sim$100 \kms \ for the [Fe VII] and [Fe X] lines (respectively) compared to the observed line shifts of 110 \kms \ and 494 \kms.  As explained in \S\ref{Geometry} our models do not include dust which would lead to increased terminal velocities (see \S 4).

In figure \ref{EmVel_19_6} we show how the \fhil \ luminosities vary as a function of velocity through an emitting cloud at $R$=10$^{19}$ cm and $n_H$=10$^6$ cm$^{-3}$.  This shows that if a cloud were to undergo acceleration from 10 \kms \ in this region of the parameter space, the [Fe X] line would be overproduced relative to the [Fe VII] line at low velocities and the [Fe VII] line would be observed at \textit{higher} blueshifts.  This is the opposite of what is seen in the spectra (see fig. \ref{Profiles}) and implies that the \fhil \ emitting clouds \textit{cannot} be undergoing acceleration in the regions of the parameter space at which these lines are produced at the observed $k$-ratios.  Instead, they must have obtained their observed velocities prior to reaching these regions.

Although regions of the parameter space with log($U)=-0.4$ can explain the low blueshift [Fe X] component and the observed $k$-ratios, it cannot account for the higher velocity [Fe X] component.  A possible source of this emission are regions of the parameter space with log($U$)=0.1 (see fig. \ref{EmVel_17_95}).  Here the luminosity of the [Fe X] and [Fe XI] lines peak at velocity $v\sim$160 km s$^{-1}$.  However, the integrated [Fe X] and [Fe XI] flux at $v\leq$110 km s$^{-1}$ correspond to 35\% and 70\% (respectively) of the total line flux.  We would therefore expect to see significant flux at low blueshifts even at this higher value of the ionisation parameter where the clouds are most strongly driven. 

In placing limits on the [O III] flux (relative to [Fe VII] flux) we are able to derive upper limits on the thickness, $\Delta R$, of the clouds before this line dominates (see \S\ref{Dynamics}).  We find that in the case of the [Fe VII] emitting clouds with log($U)=-0.4$, $\Delta R$ is typically $<$ 1\% of $R$, corresponding to a column density $N_H\sim10^{22}$ cm$^{-2}$.  Prior to reaching this depth, the [Fe VII] emissivity falls by two orders of magnitude over the final 70\% of the cloud.  This implies that no \fhil \ flux is lost in introducing the [O III]/[Fe VII] $>$6.2 stopping criteria.  In the case of the clouds described in the previous paragraph (with log($U$)=0.1) the models are stopped when the column density limit ($10^{24}$ cm$^{-2}$) is reached.  As expected, the emissivity of the [Fe X] line drops significantly before this depth is reached.

\section{Discussion}
\label{Discussion}
The results from the models described here indicate that both SEDs are capable of producing the observed $k$-ratios along a narrow range of ionisation parameters. This implies that the blue wing of the [O III] emission in Ark 564 can be produced by the same clouds as are responsible for the [Fe VII] line and the low velocity [Fe X] and [Fe XI] components.  However, these models also indicate that neither SED is capable of producing sufficient radiative driving in the regions of the parameter space where the \fhils \  are most strongly produced.  This suggests that an alternative launching mechanism is required to accelerate the \fhil \ emitting clouds to the velocities seen in these lines' blueshifts.

In the previous section we showed that the region responsible for the observed $k$-ratios (log($U)=-0.4$) could not be undergoing significant acceleration, otherwise we would see the effects of this acceleration in the profiles of the lines; there would be more [Fe~X] flux than [Fe VII] flux at low blueshifts.  These clouds must be accelerated to the observed velocities \textit{prior} to emitting \fhils.  A possible location for this accelerating process is in the previously mentioned [Fe~X] and [Fe XI] emitting clouds with log($U$)=0.1.  However, despite the [Fe X] and [Fe XI] flux peaking at $v\sim$ 160 km s$^{-1}$ in these clouds, a significant proportion of the total line flux will, again, be produced at lower velocities during the acceleration stage.  Furthermore, our models suggest that even at these increased ionisation parameters, radiative driving still has difficulty in producing the observed line shift velocities ($\sim$500 km s$^{-1}$ for [Fe X] and [Fe XI]).  

Our results indicate that \textit{potential} \fhil \ emitting clouds undergo acceleration to the observed velocities prior to these lines being produced.  Post acceleration, they then move outward to larger radii where the conditions are suitable for \fhil \ production (e.g. the torus winds described in \citealt{Dorodnitsyn08}).

In fig. \ref{Contours} we show that radiative driving is capable of producing velocities of the order of those seen in the blueshifts of the [Fe X] and [Fe XI] lines at $10^{17} < R/\textrm{cm} < 10^{18}$, corresponding to 0.03 $-$ 0.3 pc.  At these radii the energy-density temperature, which is roughly the temperature grains would assume, is in the range from 800 K to 2500 K for SED-X.  This straddles the dust sublimation temperature ($\sim$1500 K, \citealt{Barvainis87}) and is therefore the region at which the inner edge of the putative dusty torus would lie.  However, some of the key \fhil \ emitting elements (and in particular iron) are heavily depleted in dusty regions so the grains must have been destroyed by the time the gas becomes highly ionized. 

We therefore suggest that the illuminated face of the torus is the origin.   We expect dust to have a range of grain sizes and a corresponding range of temperatures, with smaller dust grains being hotter than larger grains.  These size and temperature distributions modify the concept of a single uniform dust sublimation radius, replacing it with a range of radii at which different size dust grains are evaporated. Further, the AGN luminosity is variable so the position where a given dust temperature occurs changes in response.  The large opacity of dusty gas, typically 3-4 dex above Thomson, means that the material will be radiatively driven outward while the dust is being sublimated.  In this scenario we observe the outflowing material (in the form of \fhils) when the dust extinction becomes small (\citealt{Netzer93}).  The observed coronal line region is the result.

This model facilitates the production of [Fe X] and [Fe XI] in clouds across the large range of modelled densities, yet showing similar kinematics as the terminal velocity of the clouds launched from these radii is largely independent of the density of the cloud. 

A launch from the torus in Ark 564 can account for the high velocity component of the [Fe X] and [Fe XI] lines.  Without modification, however, this model would also produce the [Fe VII] lines at these same terminal velocities.  To account for the lower velocity shift of the [Fe VII] line (and the corresponding components in [Fe X] and [Fe XI]) a retardation force must be present at $R \gtrsim 3\times 10^{18}-10^{19}$ cm ($1 - 3$ pc).  Beyond this radius our models indicate that more than 80\% of the [Fe VII] flux, 37\% of the [Fe X] flux and 35\% of the [Fe XI] flux is produced, which is in rough agreement with the observed ratios given in table \ref{Components}.  

We note that the stellar bulge of the host galaxy becomes an important factor at these large radii.  An estimate of the radius at which the stellar bulge becomes important can be determined by assuming a bulge density distribution described by \cite{Hernquist90}, a total bulge mass of $4.6\times10^{10} $M$_\odot$ (\citealt{Bian03}) and an approximate half-light radius for Ark 564 $\sim$1 kpc (determined from 2MASS K-band images; this is similar in scale to the typical half-light radii for other SBb galaxy bulges found by \cite{Kent85}).  This reveals that the gravitational potential of the stellar bulge becomes important (i.e. greater than the black hole potential) at $R=0.03$ pc, and the gravitational force of the bulge dominates the radiative+gravitational force of the black hole/accretion disk at $\sim$10 pc.  More generally \cite{Davies07} showed that there is considerable emission from young stars within the central few parsecs of an AGN-hosting galaxy.  It is thus reasonable to consider that the kinematics of the outer \fhil \ emitting region can be affected by the host galaxy at radii approaching a few parsecs from the central engine.

A number of studies have suggested that winds could be launched from the inner edge of the torus (e.g. \citealt{Krolik86}, \citeyear{Krolik88} and \citealt{Dorodnitsyn08}).  The results presented here suggest that these same winds produce \fhil \ emission once they reach the optimal ionisation parameter for their production.

\

\noindent
GJF is funded by NASA (07-ATFP07-0124) and NSF (AST0607028).

\begin{thebibliography}{}

\bibitem[\protect\citeauthoryear{{Barvainis}}{{Barvainis}}{1987}]{Barvainis87}
{Barvainis} R.,  1987, ApJ, 320, 537

\bibitem[\protect\citeauthoryear{{Bian} \& {Zhao}}{{Bian} \&
  {Zhao}}{2003}]{Bian03}
{Bian} W.-H.,  {Zhao} Y.-H.,  2003, PASJ, 55, 143

\bibitem[\protect\citeauthoryear{{Botte}, {Ciroi}, {Rafanelli} \& {Di
  Mille}}{{Botte} et~al.}{2004}]{Botte04}
{Botte} V.,  {Ciroi} S.,  {Rafanelli} P.,    {Di Mille} F.,  2004, AJ, 127,
  3168

\bibitem[\protect\citeauthoryear{{Davies}, {Mueller S{\'a}nchez}, {Genzel},
  {Tacconi}, {Hicks}, {Friedrich} \& {Sternberg}}{{Davies}
  et~al.}{2007}]{Davies07}
{Davies} R.~I.,  {Mueller S{\'a}nchez} F.,  {Genzel} R.,  {Tacconi} L.~J.,
  {Hicks} E.~K.~S.,  {Friedrich} S.,    {Sternberg} A.,  2007, ApJ, 671, 1388

\bibitem[\protect\citeauthoryear{{De Robertis} \& {Osterbrock}}{{De Robertis}
  \& {Osterbrock}}{1984}]{DeRobertis84}
{De Robertis} M.~M.,  {Osterbrock} D.~E.,  1984, ApJ, 286, 171

\bibitem[\protect\citeauthoryear{{Dorodnitsyn}, {Kallman} \&
  {Proga}}{{Dorodnitsyn} et~al.}{2008}]{Dorodnitsyn08}
{Dorodnitsyn} A.,  {Kallman} T.,    {Proga} D.,  2008, ApJL, 675, L5

\bibitem[\protect\citeauthoryear{{Erkens}, {Appenzeller} \& {Wagner}}{{Erkens}
  et~al.}{1997}]{Erkens97}
{Erkens} U.,  {Appenzeller} I.,    {Wagner} S.,  1997, A\&A, 323, 707

\bibitem[\protect\citeauthoryear{{Ferguson}, {Korista} \& {Ferland}}{{Ferguson}
  et~al.}{1997}]{Ferguson97}
{Ferguson} J.~W.,  {Korista} K.~T.,    {Ferland} G.~J.,  1997, ApJS, 110, 287

\bibitem[\protect\citeauthoryear{{Ferland}, {Korista}, {Verner}, {Ferguson},
  {Kingdon} \& {Verner}}{{Ferland} et~al.}{1998}]{Ferland98}
{Ferland} G.~J.,  {Korista} K.~T.,  {Verner} D.~A.,  {Ferguson} J.~W.,
  {Kingdon} J.~B.,    {Verner} E.~M.,  1998, PASP, 110, 761

\bibitem[\protect\citeauthoryear{{Henney}, {Arthur}, {Williams} \&
  {Ferland}}{{Henney} et~al.}{2005}]{Henney05}
{Henney} W.~J.,  {Arthur} S.~J.,  {Williams} R.~J.~R.,    {Ferland} G.~J.,
  2005, ApJ, 621, 328

\bibitem[\protect\citeauthoryear{{Hernquist}}{{Hernquist}}{1990}]{Hernquist90}
{Hernquist} L.,  1990, ApJ, 356, 359

\bibitem[\protect\citeauthoryear{{Kent}}{{Kent}}{1985}]{Kent85}
{Kent} S.~M.,  1985, ApJS, 59, 115

\bibitem[\protect\citeauthoryear{{Krolik} \& {Begelman}}{{Krolik} \&
  {Begelman}}{1986}]{Krolik86}
{Krolik} J.~H.,  {Begelman} M.~C.,  1986, ApJL, 308, L55

\bibitem[\protect\citeauthoryear{{Krolik} \& {Begelman}}{{Krolik} \&
  {Begelman}}{1988}]{Krolik88}
{Krolik} J.~H.,  {Begelman} M.~C.,  1988, ApJ, 329, 702

\bibitem[\protect\citeauthoryear{{Landt}, {Bentz}, {Ward}, {Elvis}, {Peterson},
  {Korista} \& {Karovska}}{{Landt} et~al.}{2008}]{Landt08}
{Landt} H.,  {Bentz} M.~C.,  {Ward} M.~J.,  {Elvis} M.,  {Peterson} B.~M.,
  {Korista} K.~T.,    {Karovska} M.,  2008, ApJS, 174, 282

\bibitem[\protect\citeauthoryear{{Mueller S{\'a}nchez}, {Davies}, {Eisenhauer},
  {Tacconi}, {Genzel} \& {Sternberg}}{{Mueller S{\'a}nchez}
  et~al.}{2006}]{Mueller06}
{Mueller S{\'a}nchez} F.,  {Davies} R.~I.,  {Eisenhauer} F.,  {Tacconi} L.~J.,
  {Genzel} R.,    {Sternberg} A.,  2006, A\&A, 454, 481

\bibitem[\protect\citeauthoryear{{Mullaney} \& {Ward}}{{Mullaney} \&
  {Ward}}{2008}]{Mullaney08}
{Mullaney} J.~R.,  {Ward} M.~J.,  2008, MNRAS, 385, 53

\bibitem[\protect\citeauthoryear{{Netzer} \& {Laor}}{{Netzer} \&
  {Laor}}{1993}]{Netzer93}
{Netzer} H.,  {Laor} A.,  1993, ApJL, 404, L51

\bibitem[\protect\citeauthoryear{{Osterbrock}}{{Osterbrock}}{1981}]{Osterbrock%
81}
{Osterbrock} D.~E.,  1981, ApJ, 246, 696

\bibitem[\protect\citeauthoryear{{Penston}, {Fosbury}, {Boksenberg}, {Ward} \&
  {Wilson}}{{Penston} et~al.}{1984}]{Penston84}
{Penston} M.~V.,  {Fosbury} R.~A.~E.,  {Boksenberg} A.,  {Ward} M.~J.,
  {Wilson} A.~S.,  1984, MNRAS, 208, 347

\bibitem[\protect\citeauthoryear{{Riffel}, {Rodr{\'{\i}}guez-Ardila} \&
  {Pastoriza}}{{Riffel} et~al.}{2006}]{Riffel06}
{Riffel} R.,  {Rodr{\'{\i}}guez-Ardila} A.,    {Pastoriza} M.~G.,  2006, A\&A,
  457, 61

\bibitem[\protect\citeauthoryear{{Riffel}, {Storchi-Bergmann}, {Winge},
  {McGregor}, {Beck} \& {Schmitt}}{{Riffel} et~al.}{2008}]{Riffel08}
{Riffel} R.~A.,  {Storchi-Bergmann} T.,  {Winge} C.,  {McGregor} P.~J.,  {Beck}
  T.,    {Schmitt} H.,  2008, MNRAS, 385, 1129

\bibitem[\protect\citeauthoryear{{Romano et al.}}{{Romano et
  al.}}{2004}]{Romano04}
{Romano et al.} 2004, ApJ, 602, 635

\end{thebibliography}

\bsp

\label{lastpage}

\end{document}